\documentclass[journal=apchd5,manuscript=article]{achemso}

\usepackage[version=3]{mhchem} 

\usepackage{xcolor}
\usepackage{soul}

\author{Mohamed A. Mousa}
\affiliation[Purdue ECE]{Elmore Family School of Electrical and Computer Engineering, Purdue University, West Lafayette, IN 47907, USA}
\alsoaffiliation[Birck]{Birck Nanotechnology Center, Purdue University, West Lafayette, IN 47907, USA}

\author{Leif Bauer}
\affiliation[Purdue ECE]{Elmore Family School of Electrical and Computer Engineering, Purdue University, West Lafayette, IN 47907, USA}
\alsoaffiliation[Birck]{Birck Nanotechnology Center, Purdue University, West Lafayette, IN 47907, USA}

\author{Daien He}
\affiliation[Purdue ECE]{Elmore Family School of Electrical and Computer Engineering, Purdue University, West Lafayette, IN 47907, USA}
\alsoaffiliation[Birck]{Birck Nanotechnology Center, Purdue University, West Lafayette, IN 47907, USA}

\author{Sakshi Gupta}
\affiliation[Purdue Physics]{Department of Physics and Astronomy, Purdue University, West Lafayette, IN 47907, USA}

\author{Shubhankar Jape}
\affiliation[Purdue ECE]{Elmore Family School of Electrical and Computer Engineering, Purdue University, West Lafayette, IN 47907, USA}
\alsoaffiliation[Birck]{Birck Nanotechnology Center, Purdue University, West Lafayette, IN 47907, USA}

\author{Utkarsh Singh}
\affiliation[Purdue ECE]{Elmore Family School of Electrical and Computer Engineering, Purdue University, West Lafayette, IN 47907, USA}
\alsoaffiliation[Birck]{Birck Nanotechnology Center, Purdue University, West Lafayette, IN 47907, USA}

\author{Bhagwati Prasad}
\affiliation[IISc]{Department of Materials Engineering, Indian Institute of Science, Bangalore, KA, 560012, India}

\author{Partha P. Mukherjee}
\affiliation[Purdue ME]{School of Mechanical Engineering, Purdue University, West Lafayette, IN 47907, USA}

\author{Angshuman Deka}
\affiliation[Birck]{Birck Nanotechnology Center, Purdue University, West Lafayette, IN 47907, USA}

\author{Zubin Jacob}
\email{zjacob@purdue.edu}
\affiliation[Purdue ECE]{Elmore Family School of Electrical and Computer Engineering, Purdue University, West Lafayette, IN 47907, USA}
\alsoaffiliation[Birck]{Birck Nanotechnology Center, Purdue University, West Lafayette, IN 47907, USA}
\alsoaffiliation[Purdue Physics]{Department of Physics and Astronomy, Purdue University, West Lafayette, IN 47907, USA}

\title[An \textsf{achemso} demo]
  {Ultra-broadband Mid to Long-wave Infrared Spintronic Poisson Bolometer}

\keywords{Ultra-broadband IR Sensing, Long-wave Infrared Detectors, Poisson Bolometer, Spintronics, Nanoplasmonics, Room-temperature Detector}

\begin{document}

\begin{tocentry}
  \centering
  \includegraphics[width=3.25in]{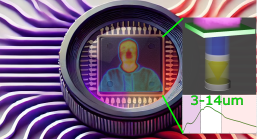}

  An uncooled spintronic Poisson bolometer leverages sub-wavelength plasmonic confinement for ultra-broadband MWIR–LWIR sensing, enabling a Poisson-counting paradigm for uncooled thermal imaging.
\end{tocentry}

\begin{abstract}
Infrared detectors have traditionally been divided into two fundamental classes, mid-wave (MWIR, 3-5 µm) and long-wave (LWIR, 8-14 µm). Integrating MWIR and LWIR within a single device is challenging due to distinct materials, cooling needs, and detection mechanisms, while such integration is critical for improved object recognition, temperature estimation, and environmental sensing. In this work, we demonstrate a Spintronic Poisson (SP) bolometer enabling room-temperature ultra-broadband sensing across 3-14 µm. Unlike conventional bolometers that rely on continuous analog signals, the SP bolometer implements a Poisson-counting detection paradigm, encoding temperature in discrete stochastic events, which turns thermal noise from a limitation into the basis of the estimator itself. We fabricate the SP bolometer using a spintronic transduction layer integrated with a plasmonic nanoantenna array to enhance broadband infrared absorption. Using spintronic transduction, the device achieves the noise-equivalent temperature difference (NETD, thermal sensitivity metric) of $\approx$80–100 mK at 300 K, surpassing uncooled detectors and approaching cooled technologies. This work establishes a statistical detection paradigm for room-temperature infrared sensing with broad application potential.
\end{abstract}

\section*{Introduction}

Thermal infrared (IR) detection for focal-plane array (FPA) imaging is primarily limited to two detector classes, mid‑wave (MWIR, 3–5µm) and long‑wave (LWIR, 8–14µm). MWIR photodetectors based on HgCdTe or InSb typically require cryogenic cooling to suppress dark current and achieve high sensitivity \cite{rogalski2012progress,razeghi2014advances,rogalski2003infrared,rogalski2003hgcdte,zha2022infrared,hafiz2019silver}. Many LWIR imagers use uncooled microbolometer focal-plane arrays based on vanadium-oxide ($VO_x$) or amorphous silicon thermistor films \cite{rogalski2023infrared,wang2013nanostructured,yadav2022advancements,das2023thermodynamically}. These two technologies rely on fundamentally different material systems, such as narrow‑bandgap semiconductors versus $VO_x$-based resistive films, and different detection mechanisms. These differences have historically prevented coherent integration of the two approaches within a single device, limiting the development of ultra-broadband infrared imaging arrays necessary for next-generation AI-driven remote sensing \cite{kazanskiy2025comprehensive} and environmental monitoring \cite{dong2025advancements}.

This broadband detection enables precise temperature estimation \cite{hagenaars2020single}, reliable material differentiation \cite{tan2020non,jang2016experimental}, and robust target identification in complex environments \cite{zhu2021multispectral,qin2023whole}. Capturing information across multiple bands provides complementary insights. MWIR offers high-contrast thermal imaging, particularly for high-temperature targets \cite{wilson2015review,rogalski2021trends}, and LWIR excels at resolving fine temperature differences in or near ambient-temperature scenes \cite{huang2024broadband}. These combined capabilities are essential for precise thermal object-detection \cite{wang2024spinning}, neural thermal imaging \cite{boiko2022integration}, predictive diagnostics such as early detection of solid-state battery failures \cite{hatzell2020challenges}, gas detection \cite{yang2023ch4}, and heat-assisted detection and ranging (HADAR) \cite{bao2023heat}.

Despite significant advances, achieving simultaneous MWIR–LWIR detection within a single compact, high-performance pixel device targeting imaging with FPAs remains a critical challenge. Conventional dual-band MWIR–LWIR sensors rely on heterogeneous stacks or hybrid optical paths, introducing cross-talk and transmission losses \cite{vincent2015fundamentals,macleod2010thin}. Lattice and bandgap mismatches between MWIR and LWIR materials hinder monolithic integration and reduce quantum efficiency \cite{rogalski2003infrared}. At the system level, most high-sensitivity MWIR detectors require cryogenic cooling, whereas LWIR detectors operate uncooled, creating noise mismatches and increasing size, weight, and power (SWaP) demands \cite{rogalski2019infrared}. These limitations prevent the realization of ultra-broadband (3–14 µm) systems that combine high thermal sensitivity ($<100 mK$), high speed ($>50 Hz$), and room-temperature (300 K) operation on a single chip.

\begin{figure*}[h!] 
\centering 
\includegraphics[width=1\textwidth]{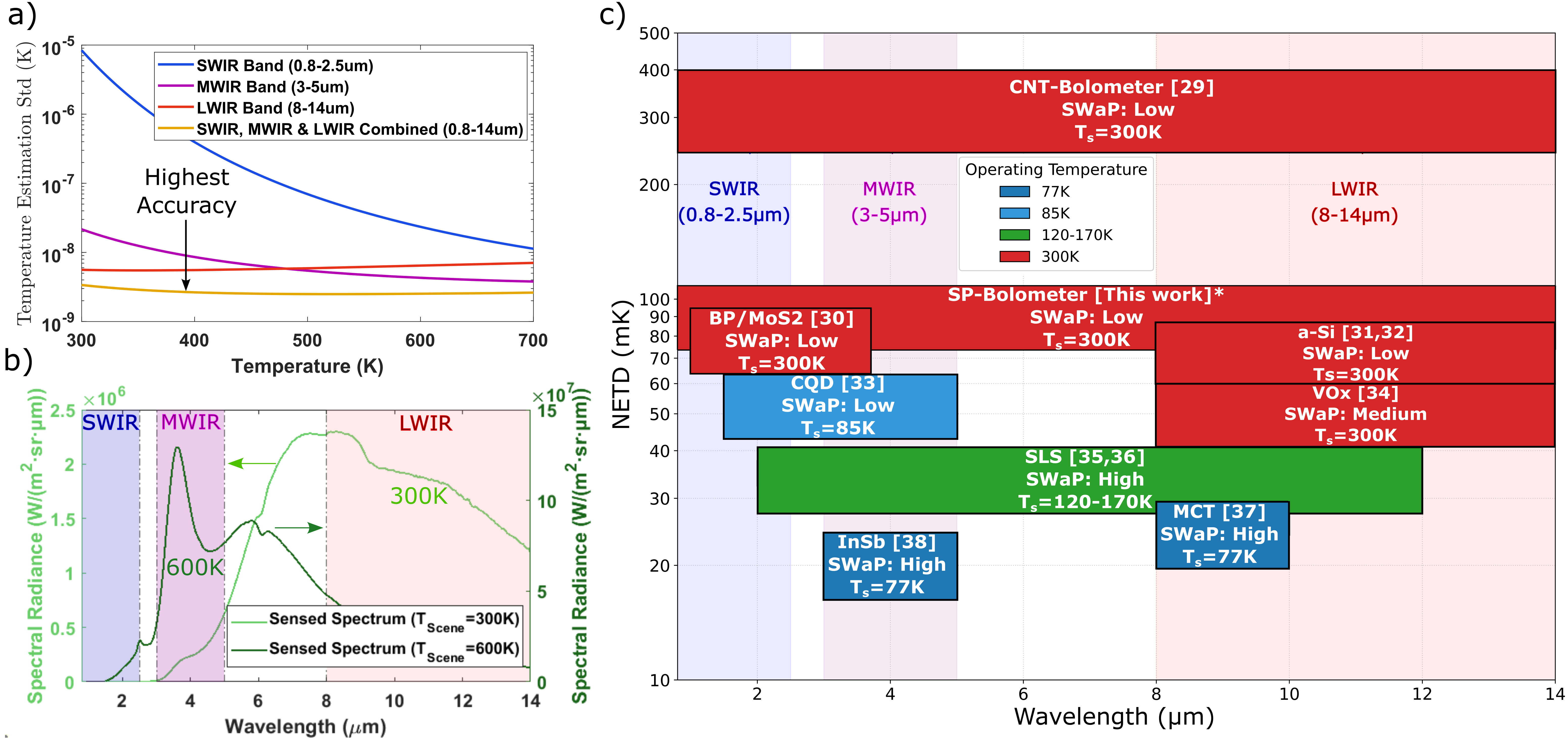} 
\caption{Ultra-broadband SP-bolometer performance and comparison with state-of-the-art infrared detectors. (a) Temperature estimation standard deviation (square root of Cramer-Rao bound (CRB)) for the SWIR, MWIR, LWIR, and the SWIR-MWIR-LWIR combined cases. Ultra-broadband combined achieves the lowest CRB (i.e., the lowest standard deviation), indicating superior accuracy compared to single-band configurations. (b) Ultra-broadband sensed spectrum of the SP-bolometer, where light green is at a scene temperature of 300K, and dark green is at a scene temperature of 600K. (c) Performance comparison of infrared detectors showing noise-equivalent temperature difference (NETD) versus wavelength for different technologies. Lower NETD means higher thermal sensitivity. Colored bars indicate operating temperatures (see legend), with labels showing detector name, SWaP, and Sensor temperature. SWaP: size, weight, and power. The Spintronic Poisson bolometer (SP, this work) achieves room-temperature operation across 0.8-14 $\mu m$ with 80-100 mK NETD. CNT: Carbon nanotube \cite{liu2018high}, BP/$MoS_2$: Black Phosphorus on top of Molybdenum Disulfide \protect\cite{shu2024high}, a-Si: Amorphous Silicon \cite{rogalski2016challenges,lynredAthena1920}, CQD: colloidal quantum dots \protect\cite{tang2019dual}, VOx: Vanadium Oxide \protect\cite{lynredPico1024Gen2}, SLS: Strained Layer Superlattice \protect\cite{lee2023design,irc906sls}, MCT: Mercury Cadmium Telluride \protect\cite{leonardoCondorHD}, InSb: Indium Antimonide \protect\cite{flirA6750sc}. *MWIR and LWIR are experimentally characterized in this work, while SWIR performance is estimated.} 
\label{fig:CRB} 
\end{figure*}

Recent efforts to achieve ultra-broadband infrared detection have explored both mature and emerging material technologies. Conventional detectors based on HgCdTe (MCT) \cite{figgemeier2018high,cervera2015low} and type-II superlattices \cite{haddadi2017bias} offer high performance but are limited by complex fabrication, low yield, high SWaP, and cooling requirements. Strained layer superlattice (SLS) architectures \cite{lee2023design} have demonstrated dual-band MWIR/LWIR detection with improved uniformity and reduced dark current, however they still require complex epitaxial growth and cooling requirements (120-170K). Stacked colloidal quantum dots (CQDs) \cite{tang2019dual} achieve tunable SWIR–MWIR dual-band detection through size-controlled PbS and PbSe layers, providing spectral tuning and reduced crosstalk, but remain constrained by fabrication complexity and cryogenic cooling (85K). Carbon nanotube (CNT) films offer broadband absorption from the ultraviolet to the terahertz range \cite{liu2018high}, but low thermal sensitivity and integration challenges have limited their use in high-resolution imaging. Graphene-based LWIR detectors have demonstrated room-temperature operation with fast response and dynamic spectral tunability across 8--12~$\mu$m via electrostatic gating \cite{guo2024spectrally}, but remain constrained to the LWIR band.

Beyond material and architectural advances, a fundamental limitation of conventional infrared bolometers lies in the detection mechanism itself. Traditional bolometric sensors infer temperature changes from continuous analog readouts such as resistance, voltage, or current, whose estimation accuracy at room temperature is fundamentally constrained by thermal fluctuation noise \cite{richards1994bolometers}, electronic readout noise \cite{martini2025uncooled}, and low-frequency noise \cite{bielecki2024review}. Here, we introduce the Poisson bolometer, a distinct regime of bolometric operation in which infrared-induced temperature changes are estimated from variations in the mean rate of discrete output events generated by an intrinsically stochastic physical process \cite{shynk2012probability,van2004detection}. This approach is fundamentally different from photon counting. The detected events do not correspond to individual photons, but arise from thermally driven dynamics whose statistics follow a Poisson distribution \cite{kingman1992poisson,van1992stochastic}. 

In this model, the event generation follows a Poisson process characterized by a temperature-dependent rate. The probability of detecting $N$ events within a fixed measurement interval $t_{int}$ is given by the Poisson probability mass function:

\begin{equation}
P(N; \lambda_0, t_{int}) = \frac{(\lambda_0 t_{int})^N e^{-\lambda_0 t_{int}}}{N!},
\end{equation}
where $\lambda_0$ represents the baseline mean event rate (events per second) at room temperature. Under blackbody illumination of intensity $I_{\mathrm{BB}}$, the probability becomes:

\begin{equation}
P(N; \lambda_{\mathrm{BB}}, t_{int}) = \frac{(\lambda_{\mathrm{BB}} t_{int})^N 
e^{-\lambda_{\mathrm{BB}} t_{int}}}{N!}.
\end{equation}
Here, $\lambda_{\mathrm{BB}}$ is the total mean event rate, which includes the contribution from absorbed infrared power and scales linearly with $I_{\mathrm{BB}}$ \cite{pinsky2010introduction}. This establishes a statistical framework in which temperature is estimated directly from the mean frequency of stochastic events over the integration time $t_{int}$.

By encoding temperature information in discrete event counts rather than continuous signal amplitudes, the Poisson bolometer inherently operates under stochastic fluctuations present at room temperature. These fluctuations increase or decrease depending on how the detected temperature deviates from room temperature. In this regime, thermal noise is not a limitation but forms the basis of the estimator itself \cite{lehmann1998theory,van2004detection}. Importantly, this Poisson-counting regime is independent of any specific material system or device architecture and can be realized by a broad class of bolometric structures that support Poisson-distributed event generation.

One realization of the Poisson-bolometer operating regime is achieved using spintronic materials \cite{bauer2025exploiting}. In this platform, infrared-induced photothermal heating modulates stochastic magnetic switching processes, enabling direct transduction into discrete electrical events with Poisson statistics. Leveraging this operating principle, we implement the Spintronic Poisson (SP) bolometer, which couples spintronic transduction with engineered optical absorption to overcome the bandgap-limited spectral response of conventional infrared detectors. This approach decouples spectral response from material bandgap constraints, enabling continuous MWIR–LWIR sensing within a single nanoscale device without the need for cryogenic cooling and, building on this principle, achieves ultra-broadband sensitivity across 3–14 µm with measured thermal sensitivity of 80–100 mK at room temperature. The core scientific advance lies in coupling Poisson-based spintronic readout with engineered optical absorption, enabling broadband spectral coupling, high-speed operation, and compatibility with chip-scale integration. 

Building on this principle, the SP bolometer achieves ultra-broadband sensitivity across 3–14 µm with measured thermal sensitivity of 80–100 mK at room temperature. A nanoplasmonic antenna array positioned above the active region enhances infrared absorption and provides tunable spectral selectivity through controlled geometry and material design. The $200 nm \times 250 nm$ active area enables sub-wavelength pixel footprints ($< 10\mu m$), which are difficult to achieve in conventional MWIR and LWIR technologies limited by the thermal capacitance of the microbolometer element and its thermal time constant \cite{yu2020low}. Smaller pixels reduce responsivity due to reduced absorber area and limited heat capacity \cite{jung2022improved}. Figure \ref{fig:CRB}a demonstrates through a Cramér–Rao-bound analysis that jointly exploiting SWIR, MWIR, and LWIR measurements substantially improves temperature-estimation accuracy because each spectral band contributes complementary information (see Supplementary Section 1), and Figure \ref{fig:CRB}b shows the resulting broadband absorption response under 300 K and 600 K blackbody illumination.

\begin{figure*}[htbp]
    \centering
    \includegraphics[width=\textwidth]{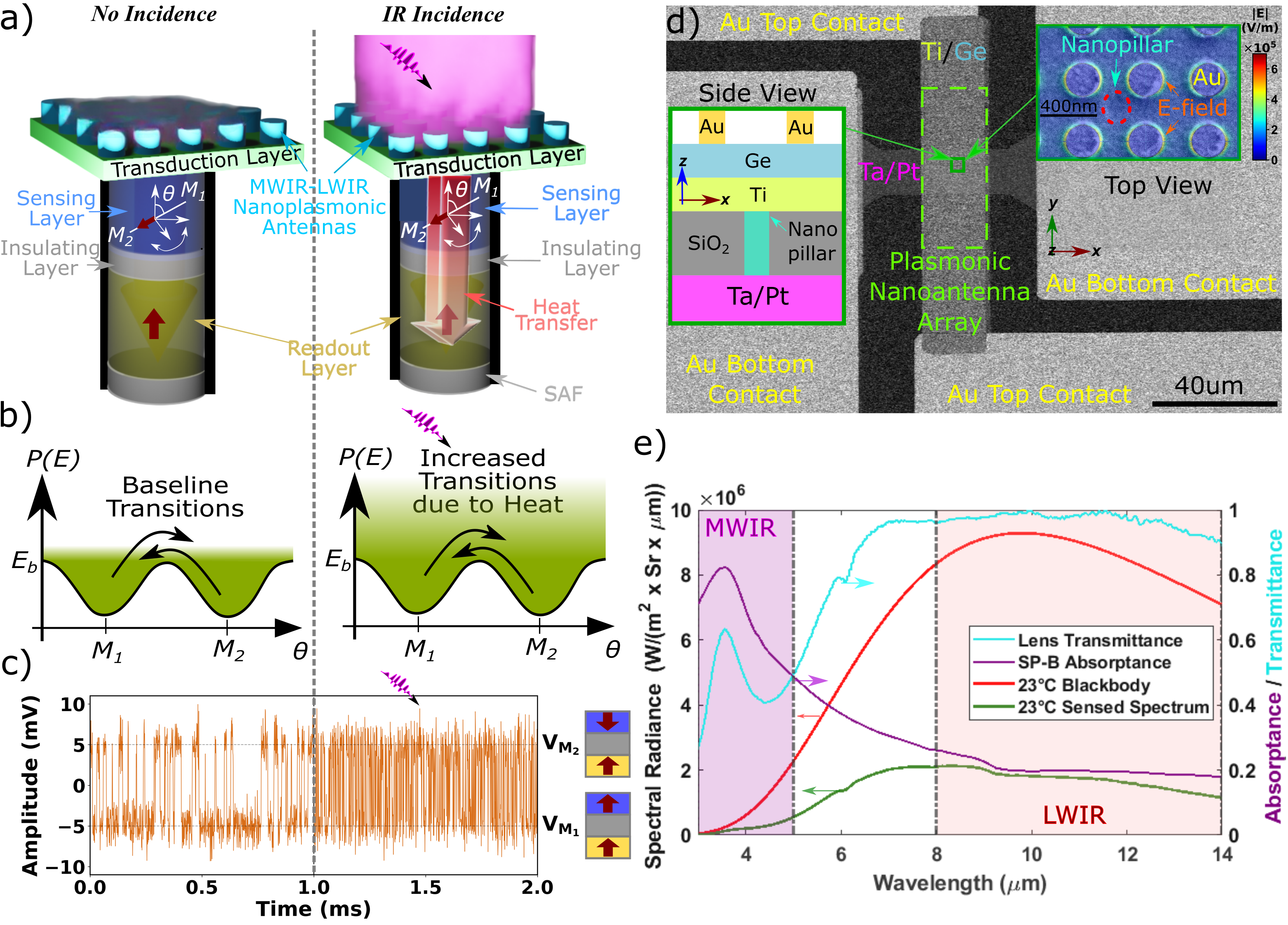}
    \caption{Schematic of SP-bolometer operation and spectral response. (a) Schematic comparing device states with no incidence (left) and IR incidence (right), where absorbed light creates a hotspot in the transduction layer that thermalizes through the device. A synthetic antiferromagnet (SAF) adjacent to the sensing layer stabilizes the magnetic orientation. (b) Energy diagram showing baseline stochastic transitions (left) and increased switching probability between magnetization states $M_1$ and $M_2$ due to absorbed heat (right). (c) The readout is obtained via resistance changes in the sensing layer, recorded as voltage transitions between two discrete levels: $V_{M_1}$ (parallel state: $\mathbf{M}_{sensing} = \mathbf{M}_{readout}$) and $V_{M_2}$ (anti-parallel state: $\mathbf{M}_{sensing} = -\mathbf{M}_{readout}$). (d) SEM image of the SP-bolometer device with a plasmonic nanoantenna array atop a transduction layer (Ge/Ti). The inset on the right shows a top view of the nanoantenna array atop the SP-bolometer nanopillar, overlaid with a COMSOL-simulated electric field distribution. The left inset shows a side view of the SP-bolometer. (e) Spectral absorptance (purple), lens transmittance (cyan), blackbody radiation (red), and sensed spectrum (green) at a scene temperature of 23 °C. The nanoantenna array provides strong MWIR absorption and significant LWIR absorption, supporting broadband sensing despite the blackbody distribution.}
    \label{fig:SP}
\end{figure*}

To benchmark the performance of the SP bolometer, we conducted a comprehensive survey of representative infrared detectors spanning the short-wave, mid-wave to long-wave spectral regions (SWIR–MWIR–LWIR), including both room-temperature and cooled technologies. Figure \ref{fig:CRB}c compares the noise-equivalent temperature difference (NETD) versus wavelength for the surveyed detectors. NETD is the key metric for thermal sensitivity in thermal imaging systems (FPAs) under blackbody scene illumination \cite{rogalski2023infrared}. The SP bolometer is characterized with a broadband blackbody source. Thus, its performance is evaluated using NETD rather than noise-equivalent power (NEP) or specific detectivity ($D^*$), which is typically defined for laser or narrowband illumination. Moreover, NEP primarily includes intrinsic dark noise, whereas NETD evaluates practical imaging performance by accounting for both device dark noise and background photon noise from the thermal scene \cite{rogalski2023infrared}. The SP bolometer reported here operates at room temperature across 3–14 µm with an NETD of 80–100 mK, establishing ultra-broadband thermal sensing performance. It outperforms state-of-the-art room-temperature detectors, including CNT (Carbon Nanotube) \cite{liu2018high}, BP/MoS$_2$ junction FET photodetectors (1.55–3.6 µm) \cite{shu2024high}, a-Si (Amorphous Silicon) LWIR detector \cite{rogalski2016challenges,lynredAthena1920}, and VOx (Vanadium Oxide) LWIR detector \cite{lynredPico1024Gen2}. Its performance is also comparable to leading cooled detectors, such as CQD (Colloidal Quantum Dot) detectors for NIR–MWIR with tunable band selection \cite{tang2019dual}, SLS (Strained Layer Superlattice) \cite{irc906sls,lee2023design}, MCT (Mercury Cadmium Telluride) \cite{leonardoCondorHD}, and InSb (Indium Antimonide) \cite{flirA6750sc} across their respective spectral bands (see Supplementary Section 2).

\section*{\label{Design}Device Concept and Spintronic Transduction}
We propose an ultra-broadband bolometer that integrates high-speed readout (30--120 Hz, fundamentally defined by the statistical integration time $t_{int}$ (see Supplementary section 7)), high thermal sensitivity (80-100mK) sensing with low power consumption ($<1mW$) and a compact active area (0.09µ$m^2$). This solution spans the full infrared range from MWIR to LWIR, using the spintronic Poisson bolometer and a high-speed filter wheel to enable efficient spectral demonstration.

\subsection*{\label{SPB}Stochastic Switching and Poisson Detection Regime}

To physically realize this Poisson detection regime at room temperature, the SP-bolometer is constructed using a stochastic magnetic tunnel junction (MTJ) architecture consisting of a free magnetic sensing layer with an out-of-plane magnetic easy axis, a nonmagnetic insulating tunnel barrier, and a fixed magnetic readout layer, whose orientation is stabilized by an adjacent synthetic antiferromagnet (SAF), ensuring a reliable reference magnetization state \cite{butler2001spin,mathon2001theory,singh2024cmos,aadit2022massively,safranski2021demonstration}. In this architecture, charge carriers tunnel through the thin MgO barrier (1~nm) \cite{ikeda2008tunnel,djayaprawira2005230}. Through the tunnel magnetoresistance (TMR) effect, the relative magnetic orientation of the free and fixed layers dictates the junction's resistance, yielding a low resistance in the parallel state ($M_1$: $\mathbf{M}_{sensing} = \mathbf{M}_{readout}$) and a high resistance in the anti-parallel state ($M_2$: $\mathbf{M}_{sensing} = -\mathbf{M}_{readout}$). A small DC bias ($\approx$500$\mu$A) applied across the junction converts these resistance states into measurable voltage levels, enabling direct electrical readout of the magnetic state. As shown in the schematic comparison of the device states under No Incidence and IR Incidence (Figure \ref{fig:SP}a), the bolometer operates through thermally activated transitions between two stable magnetization states ($M_1$ and $M_2$) in the spintronic sensing layer \cite{hayakawa2021nanosecond,coffey2012thermal,neel1949theorie}. Unlike conventional MTJs designed for non-volatile memory, where large energy barriers ($>60~k_BT$) ensure long-term data retention \cite{dieny2020opportunities,zink2022review,jung2022crossbar}, the SP-bolometer free layer is intentionally engineered with a small energy barrier ($E_b \approx 90$~meV, or $\approx 3.5~k_BT$ at room temperature), placing it in the thermally accessible superparamagnetic regime where room temperature thermal fluctuations drive continuous, spontaneous switching between $M_1$ and $M_2$. This low barrier is governed by the Stoner-Wohlfarth model, $E_b = \frac{M_s V (H_k \pm H)}{2}$ \cite{stoner1948mechanism,soumah2025entropy,hayakawa2021nanosecond}, where $H_k$ is the effective perpendicular anisotropy field arising from the balance between the interfacial PMA and the in-plane shape anisotropy. By scaling the magnetic volume ($V$) to the nanoscale free layer and introducing an ultra-thin Tungsten (W) insertion (0.3nm), the interfacial PMA sufficiently exceeds the in-plane shape anisotropy, establishing an out-of-plane easy axis and reducing the net energy barrier to the thermally accessible superparamagnetic regime \cite{soumah2025entropy,kim2015ultrathin,mihajlovic2020thermal}. Assuming an attempt time of $\tau_0 \sim 1$~ns \cite{brown1963thermal,kanai2021theory}, this $3.5 k_B T$ barrier yields a mean magnetic dwell time of $\sim$33~ns (a $\sim$30~MHz baseline switching rate). The device exhibits a TMR ratio of approximately 98\% and a resistance-area (RA) product of 80--110~$\Omega\cdot\mu \text{m}^2$ (see Supplementary section 8 (Methods)). Full magnetic and electrical characterization, including coercivity, offset field, and RA product uniformity, is provided in Supplementary Fig.~S11. This TMR effect converts the magnetic transitions into measurable, discrete voltage fluctuations observed as random telegraph noise (Figure \ref{fig:SP}c). Upon absorbing incident infrared light in the transduction layer (Figure \ref{fig:SP}a), a thermal hotspot is formed, which propagates heat through the bolometer. This thermal energy modifies the energy distribution in the magnetic sensing layer, exponentially increasing the thermally activated transition rate between magnetic states according to the Néel-Arrhenius law \cite{coffey2012thermal,neel1949theorie,brown1963thermal} as illustrated by the energy diagram (Figure \ref{fig:SP}b), where IR incidence results in absorbed energy that increases the switching probability \cite{hayakawa2021nanosecond}: 

\begin{equation}
\tau = \tau_0 \exp\left(\frac{E_b}{k_B T}\right)
\end{equation}
where $\tau$ is the dwell time between transitions, $\tau_0$ is the attempt time, $k_B$ is the Boltzmann constant, and $T$ is the temperature of the sensing layer \cite{brown1963thermal,kanai2021theory}.

Thus, the incident infrared heat increases the stochastic switching rate between the magnetic states ($M_1$ and $M_2$). This switching is measured via voltage transitions (Figure \ref{fig:SP}c) between two discrete levels: $V_{M_1}$ (parallel state: $\mathbf{M}_{sensing} = \mathbf{M}_{readout}$) and $V_{M_2}$ (anti-parallel state: $\mathbf{M}_{sensing} = -\mathbf{M}_{readout}$). By measuring this random telegraph noise signal in the time domain (see Supplementary section 8 (Methods)), we estimate number of the discrete switching events over a fixed integration time ($t_{int}=33ms$). This allows us to extract the temperature-dependent event rate $\lambda$ directly, linking the physical readout of the MTJ to the Poisson statistical framework established in Equations (1) and (2). Consequently, the measurement relies on a discrete counting process rather than on integrating an analog average of the device resistance. These changes are read out as discrete digital events at up to 120 Hz (see Supplementary Section 7). Figure \ref{fig:SP}d shows the SEM image of the SP-bolometer device with a transduction layer modified with a plasmonic nanoantenna array. The inset of Figure \ref{fig:SP}d shows an SEM image of the nanoantenna array on the SP bolometer nanopillar with an elliptical cross-section of approximately 200~nm $\times$ 250~nm, overlaid with the electric-field distribution, which highlights field enhancement around the plasmonic nanoantennas. 

\subsection*{\label{Nanoantenna}Plasmonic Enhancement for Broadband Absorption}

Integrating the nanoplasmonic antenna array into the SP bolometer enhances infrared absorption through localized plasmon resonance \cite{su2023boosting,debu2018tuning,tan2020non,yang2025optical}. The antennas enhance the electromagnetic field around the circular perimeter of the nanoantenna (inset of Figure \ref{fig:SP}d), leading to enhanced absorption of incident infrared radiation and improved overall device thermal sensitivity (see Supplementary Section 2). COMSOL simulation shows the field enhancement due to the nanoplasmonic antenna (Supplementary Figure 2). Our spectral measurements align well with COMSOL simulations, further validating the antenna design’s effectiveness across the MWIR-LWIR spectrum (Supplementary Figure 3). This field enhancement is critical for achieving consistent performance across the full IR spectrum. 

Figure \ref{fig:SP}e illustrates ellipsometry measurements of the nanoplasmonic antennas' absorptance (purple) peaks in the MWIR, while the blackbody spectral radiance (red) at 23°C peaks in the LWIR. The lens transmissivity (cyan) was measured by Fourier transform infrared (FTIR) spectroscopy. The design of the nanoplasmonic antenna ensures that the sensor effectively absorbs infrared radiation across the MWIR and LWIR bands, as shown in Figure \ref{fig:SP}e (green). The antenna design incorporates a transduction layer comprising 70 nm titanium (Ti) and 95 nm germanium (Ge), along with a refined array of 40 nm-thick gold (Au) plasmonic nanoantennas, each with a 300 nm diameter and 320 nm center-to-center spacing. These antennas, arranged in a square lattice, optimize interaction with infrared radiation, providing robust ultra-broadband IR sensing (Supplementary Figure 4). In addition, the absorptance peak of the nanoplasmonic antenna can be tuned by adjusting the antenna design, such as its material composition and dimensions (Supplementary Figure 5). This flexibility allows the ultra-broadband SP bolometer to target specific IR applications with characteristic absorption spectra, such as gas sensing \cite{krzempek2019review}, temperature estimation \cite{gabriel2018infrared}, and medical diagnostics \cite{lahiri2012medical,mousa2021toward}.

\begin{figure}[h!]
    \centering
    \includegraphics[width=0.51\textwidth]{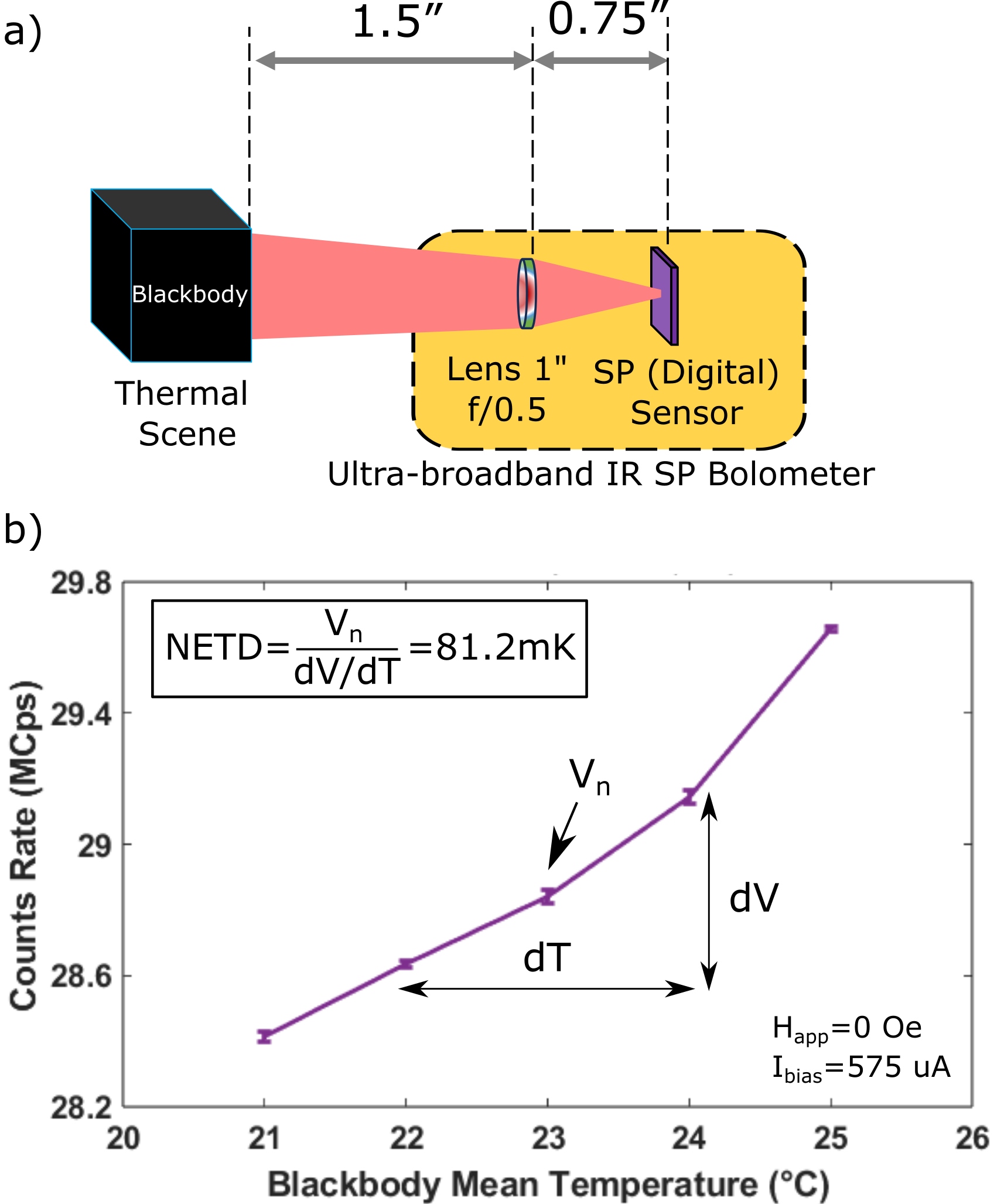}
    \caption{Broadband NETD measurement setup and response of the SP-bolometer. (a) Schematic of the broadband NETD measurement setup for the SP-bolometer, including ultra-stable blackbody source and broadband ZnSe lens (f/0.5). (b) Broadband NETD response of the SP-bolometer without optical filtering, measured at varying blackbody temperatures, demonstrating a NETD of approximately 81mK under 0Oe magnetic field and 500µA bias.}
    \label{fig:BB_NEDT}
\end{figure}

\begin{figure*}[h!]
    \centering
    \includegraphics[width=1\textwidth]{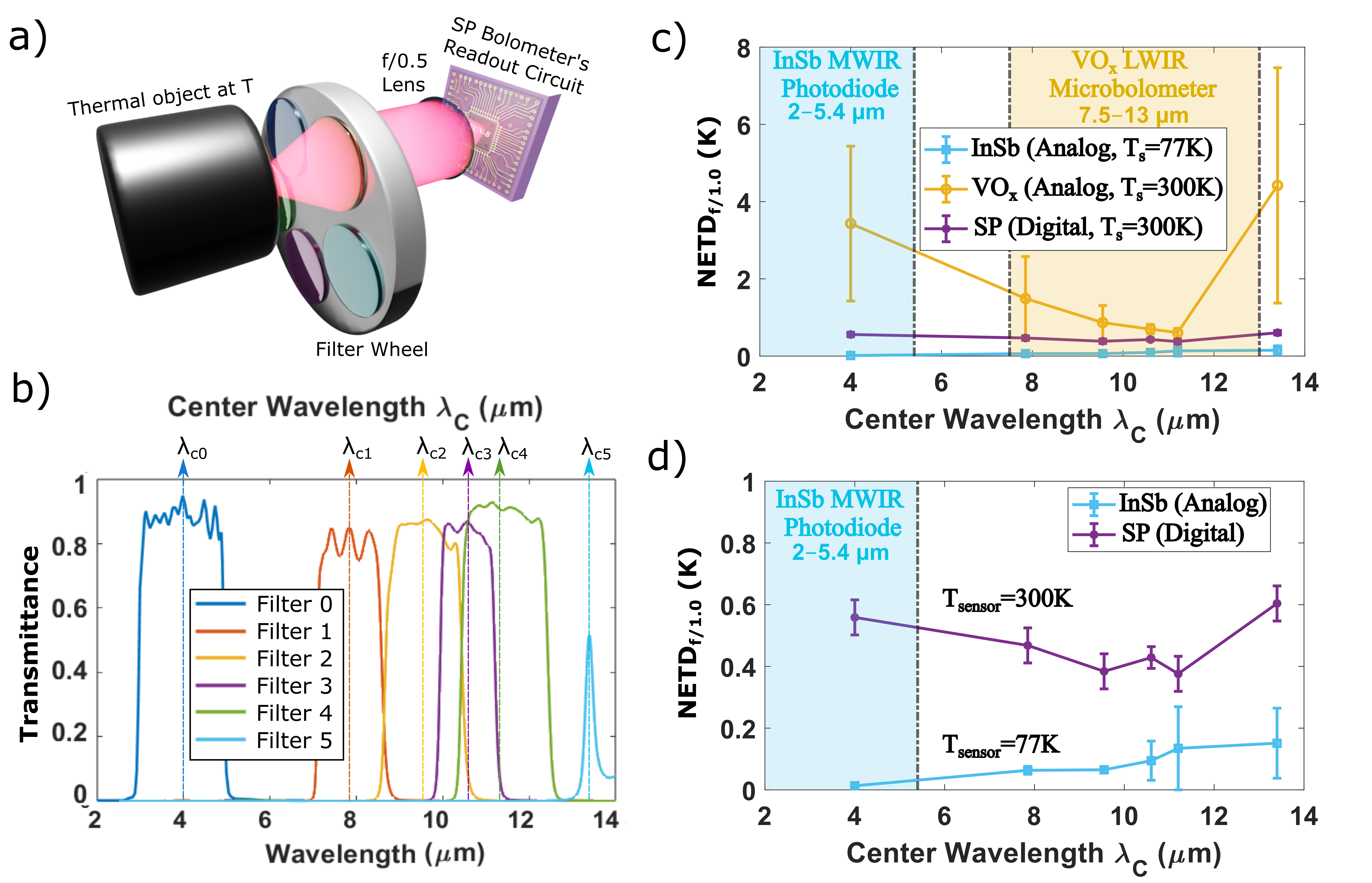}
    \caption{Consistent MWIR–LWIR NETD performance of the SP-bolometer compared with state-of-the-art infrared cameras. (a) Experimental setup for ultra-broadband MWIR–LWIR NETD measurements using the Spintronic Poisson (SP) bolometer with a high-speed filter wheel. (b) Spectral transmittance of the optical filters used in the NETD measurements. (c) Spectral comparison of NETD for a representative state-of-the-art uncooled analog $VO_x$ microbolometer (analog, LWIR, 300K, FLIR A325sc) in blue color, InSb photodiode (analog, MWIR, 77K, Telops Spark M150) in orange color, and broadband SP-bolometer (purple). The InSb photodiode performs better with lower NETD across the $3  \mu \text{m}$ to $13.5 \mu \text{m}$ range. However, the SP-bolometer's NETD is comparable to the InSb photodiode and notably outperforms the $VO_x$ microbolometer across all bands. (d) Zoomed-in view of (c), highlighting differences in NETD between the SP bolometer and InSb photodiode. Error bars for the $VO_x$ microbolometer increase at wavelengths farther from $10.6 \mu \text{m}$, likely due to the LWIR lens’s optical bandwidth limitations. $T_s$: Sensor Temperature. The NETD values are normalized to an $f$-number of $f/1.0$, where $\text{NETD}_{f/1} = \text{NETD}/(f/\#)^2$. The $f$-numbers are $f/\# = 0.5$ for the SP-bolometer, $f/\# = 0.6$ for the $VO_x$ microbolometer, and $f/\# = 2.3$ for the InSb photodiode.} 
    \label{fig:Vision}
\end{figure*}

\section*{\label{ComparativeResults} Performance Benchmarking and Thermal Sensitivity}

The thermal sensitivity of the IR detectors is characterized using NETD \cite{rogalski2019infrared}, a key metric representing the smallest detectable temperature difference in the scene. NETD is defined as:

\begin{equation}
NETD = \frac{V_n}{dV/dT}
\end{equation}

where $V_n$ is the standard deviation of the detector signal at $T_0$ (23°C in our measurements) and $dV/dT$ is the signal change per degree of temperature variation. 

Broadband NETD measurements of the SP-bolometer were conducted using an ultra-stable blackbody source ($\approx$1mK stability) and a broadband ZnSe IR lens with an f/0.5 aperture, yielding 81.2mK under 0Oe magnetic field and 500µA bias (Figure \ref{fig:BB_NEDT}a–b). Using the NETD metric, the SP-bolometer was compared to a representative LWIR $VO_x$ microbolometer (analog, 300K, FLIR A325sc) and MWIR InSb photodiode (analog, 77K, Telops Spark M150). For band-specific characterization, a high-speed MWIR–LWIR filter wheel enabled rapid selection of MWIR and LWIR bands at 30Hz, while the blackbody provided controlled thermal illumination from 21°C to 25°C in 1°C increments. The SP-bolometer signal was processed using a 33~ms moving mean window \cite{ma2023eliminating}, which serves as the statistical integration time ($t_{int}$) required to accumulate sufficient switching events for reliable Poisson rate estimation ($\mathrm{SNR} > 10~\mathrm{dB}$), defining the 30~Hz readout speed and ensuring accurate NETD measurements.

\begin{figure*}[h!]
    \centering
    \includegraphics[scale=0.48]{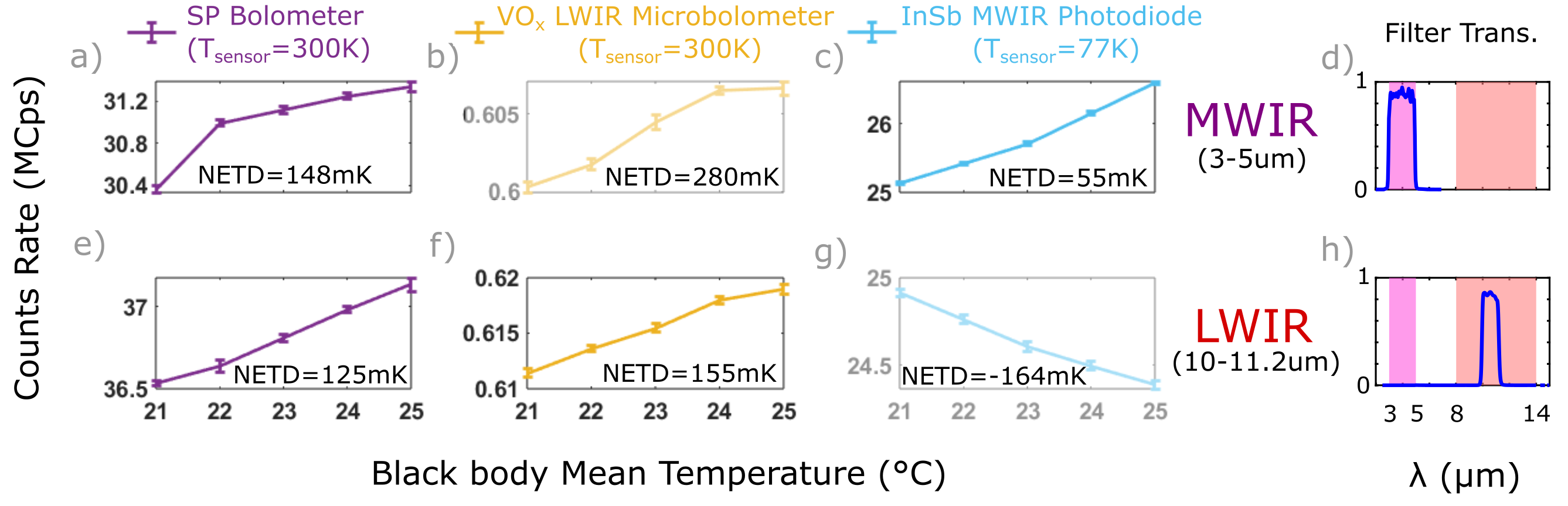}
    \caption{Consistent MWIR-LWIR sensor response of the SP-bolometer compared with state-of-the-art cameras. (a) SP-bolometer response (count rate) to the MWIR band ($3-5~\mu\text{m}$). (b) $VO_x$ microbolometer (analog, LWIR, 300K, FLIR A325sc) response (count rate) to the MWIR band ($3-5~\mu\text{m}$). (c) InSb photodiode (analog, MWIR, 77K, Telops Spark M150) response (count rate) to the MWIR band ($3-5~\mu\text{m}$). (d) MWIR filter transmittance ($3-5~\mu\text{m}$). (e) SP-bolometer response (count rate) to the LWIR band ($10-11.2~\mu\text{m}$). (f) $VO_x$ microbolometer (analog, LWIR, 300K, FLIR A325sc) response (count rate) to the LWIR band ($10-11.2~\mu\text{m}$). (g) InSb photodiode (analog, MWIR, 77K, Telops Spark M150) response (count rate) to the LWIR band ($10-11.2~\mu\text{m}$). (h) LWIR filter transmittance ($10-11.2~\mu\text{m}$). $T_s$: sensor temperature. The SP-bolometer exhibits a consistent response across MWIR and LWIR bands, performing comparably to the cooled InSb photodiode in MWIR and outperforming the $VO_x$ microbolometer across LWIR. Error bars represent the standard deviation from five measurements with 33~ms integration time.} 
    \label{fig:CR}
\end{figure*}

The experimental setup (Figure \ref{fig:Vision}a) ensured fair comparison across sensors. We used a broadband ZnSe lens (f/0.5) for the SP-bolometer, its built-in f/0.6 lens for the $VO_x$ microbolometer, while the InSb photodiode was equipped with a built-in f/2.3 lens. To account for differences in optics, NETD was normalized according to lens f-number (f/\#):

\begin{equation}
    NETD_{f/1.0} = \frac{NETD}{(f/\#)^2}
\end{equation}

Figure \ref{fig:Vision}c presents normalized NETD results for the SP-bolometer across MWIR and LWIR bands, showing performance comparable to the InSb photodiode (analog, MWIR, 77K, Telops Spark M150) and superior to the $VO_x$ microbolometer (analog, LWIR, 300K, FLIR A325sc) across all bands. Error bars indicate standard deviation over five measurements, reflecting optical bandwidth limitations for both $VO_x$ microbolometer (centered at $10.6 \mu \text{m}$, yellow shaded area) and InSb photodiode (centered at $4 \mu \text{m}$, blue shaded area).

Figure \ref{fig:CR} presents a comparative analysis of sensor responses (count rate), illustrating the SP-bolometer’s MWIR response with an MWIR filter (filter 0, Figure \ref{fig:CR}a) and LWIR response with LWIR filters (filter 3, Figure \ref{fig:CR}e). Filter transmissivity is shown in Figures \ref{fig:CR}d (MWIR) and \ref{fig:CR}h (LWIR). The SP-bolometer demonstrates performance comparable to the cooled InSb photodiode in MWIR while exceeding that of the $VO_x$ microbolometer across all bands. The InSb photodiode exhibits the strongest MWIR response but displays a monotonically decreasing count rate with increasing scene temperature in LWIR, reflecting that the detector and optics are not designed for this band.

The SP-bolometer is compatible with large-scale fabrication and system integration. Its planar, lithography-based architecture can be implemented using standard CMOS-compatible processes \cite{suhail2023first,barla2024design,singh2025long}, enabling wafer-level scalability and array-level deployment. The low-bias, room-temperature operation further supports seamless integration with on-chip readout and neuromorphic processing circuits, enabling the SP-bolometer to function simultaneously as an imaging sensor and a stochastic computing unit, achieving LWIR image classification with microsecond inference time at microwatt-level power consumption \cite{Mousa2025Neural}. The discrete, event-driven nature of the Poisson-counting readout is inherently suited for event-based infrared imaging, where only scene changes are detected rather than full frames, enabling high-speed motion detection exceeding 1~kHz in the LWIR. This approach has recently been demonstrated using the SP-bolometer platform  \cite{mousa2026uncooled}. Future work will focus on extending the operational bandwidth beyond the 0.8--14~$\mu$m range (toward the THz regime) through absorber and antenna engineering, and on integrating multispectral filtering to enable intelligent, adaptive sensing and support heat-assisted detection and ranging (HADAR) \cite{bao2023heat} applications. These developments position the SP-bolometer as a foundational platform for next-generation ultra-broadband infrared imaging and perception technologies.

\section*{\label{Conclusion} Conclusion}
This work demonstrates the design and experimental characterization of the ultra-broadband SP-bolometer, a non-cryogenic infrared detector with high thermal sensitivity across the MWIR–LWIR spectrum. The SP-bolometer achieves a broadband NETD of 81 mK under ambient operation (300 K) with minimal biasing requirements (0 Oe, 500 µA), demonstrating high thermal sensitivity. Comparative characterization shows that the SP-bolometer response is comparable to that of an InSb photodiode (analog, MWIR, 77K) and surpasses that of the $VO_x$ microbolometer (analog, LWIR, 300K) across all spectral bands. By encoding temperature in discrete Poisson-distributed events rather than continuous analog signals, the SP-bolometer operates natively under thermal fluctuations, effectively turning noise into a measurable signal and establishing a fundamentally new detection paradigm. The integrated nanoplasmonic antennas enable efficient and broadband infrared absorption, while the 33~ms moving mean integration window minimizes signal fluctuations to maintain high measurement accuracy at 30Hz. This demonstration establishes the SP-bolometer as a compact, energy-efficient, and ultra-broadband infrared sensor that advances uncooled detection technologies for applications in autonomous systems, environmental sensing, and biomedical diagnostics. Its room-temperature operation, CMOS-compatible planar design, and low-power architecture further enable scalable array integration and intelligent on-chip infrared imaging.


\section*{Funding Sources}
This work was partially supported by an Elmore Chaired Professorship at Purdue University.

\begin{acknowledgement}
We would like to thank Dr. Tiffany Santos from Western Digital for providing the thin films used to fabricate the SP bolometer devices in this work.
\end{acknowledgement}

\section*{Associated Content}

\textbf{Pre-print Statement:} Mousa, M. A.; Bauer, L.; He, D.; Gupta, S.; Jape, S.; Singh, U.; Prasad, B.; Mukherjee, P. P.; Deka, A.; Jacob, Z. Ultra-broadband Mid to Long-wave Infrared Spintronic Poisson Bolometer. 2026, arXiv:2601.11733. arXiv.org. \url{https://arxiv.org/abs/2601.11733} (accessed March 21, 2026).

\begin{suppinfo}
Theoretical modeling, COMSOL simulations, experimental setups, spectral sensitivity data, SNR analysis, and fabrication methods (PDF)
\end{suppinfo}
\bibliography{achemso-demo}

\end{document}